# Exploring Factors Affecting Student Learning Satisfaction during COVID-19 in South Korea

JIWON HAN

Department of Statistics, Sungkyunkwan University, jiwon.h@g.skku.edu

CHAE EUN RYU

Department of Computer Education, Sungkyunkwan University, superbunny38@gmail.com

GAYATHRI NADARAJAN

Department of Computing and Informatics (Data Science), Sungkyunkwan University, gaya@g.skku.edu

Understanding students' preferences and learning satisfaction during COVID-19 has been primarily centred on online learning and has focused on learning attributes such as self-efficacy, performance, and engagement. Although existing efforts have constructed statistical models capable of accurately identifying significant factors impacting learning satisfaction, these models do not necessarily explain the complex relationships among these factors in depth. This study aimed to understand several facets related to student learning preferences and satisfaction during the pandemic such as individual learner characteristics, instructional design elements, and social and environmental factors. This could aid institutions and educators tailor more effective teaching strategies. Responses from 302 students from Sungkyunkwan University in South Korea were collected between November 2021 and November 2022. Information gathered included their gender, study major, satisfaction and motivation levels when learning, perceived performance, emotional state and learning environment. Wilcoxon Rank sum test and Explainable Boosting Machine (EBM) were performed to determine significant differences in specific cohorts. The two core findings of the study are as follows: (1) Using Wilcoxon Rank Sum test, we can attest with 95% confidence that students who took offline classes had significantly higher learning satisfaction, among other attributes, than those who took online classes as with Science, Technology, Engineering and Medicine (STEM) versus Humanities and Social Sciences (HASS) students; (2) An explainable boosting machine (EBM) model fitted to 95.08% accuracy determined the top five factors affecting students' learning satisfaction as: their perceived performance, their perception on participating in class activities, their study majors, their ability to conduct discussions with classmates and the study space availability at home. Positive perceived performance and ability to discuss with classmates had a positive impact on learning satisfaction, while negative perception on class activities participation had a negative impact on learning satisfaction.

## 1 INTRODUCTION

The COVID-19 pandemic led to significant changes in higher education, with online learning becoming the main teaching method for over two years in South Korea. Educators had to quickly adapt to different teaching methods, such as live-streamed classes and pre-recorded lectures, often without sufficient time to optimize these for student satisfaction and motivation. Additionally, the pandemic created new concerns regarding students' emotional well-being and coping strategies. To study these aspects, we surveyed 302 students from Sungkyunkwan University in South Korea between November 2021 and 2022. The survey examined factors such as learning satisfaction, perceived performance, motivation, and coping strategies after four semesters of online learning. We present key findings, differences in satisfaction among student groups, and insights from an Explainable Boosting Machine (EBM) model that helped identify the critical factors influencing students' learning satisfaction.

The rest of the article is organized as follows. Section 2 clarifies what is learning satisfaction and how it has been defined and measured in various studies. Section 3 presents the methodology adopted for designing the survey. Section 4 puts forth statistically significant comparisons of learning attributes identified between two different groups. Section 5 contains the method and findings related to the factors influencing students' learning satisfaction Finally, Section 6 concludes the overall findings, and discusses future directions for this work.

## 2 LITERATURE REVIEW

### 2.1 Definition of learning satisfaction

Learning satisfaction has been defined in various ways throughout the literature, but there is a general consensus that it refers to the level of fulfillment of a learner's needs and desires. Flammger [1] defined satisfaction as the realization of needs, the joy of fulfillment, and the feeling of sufficiency. Venkateswarlu et al. [2] defined learning satisfaction as the state of mind a person experiences with the outcome in terms of fulfillment of the needs or desires. Similarly, learning satisfaction was defined as a feeling or attitude of learners that their desires and needs can be fulfilled in learning activities or processes ([3], [4])

### 2.2 Features related to learning satisfaction

Learning satisfaction is a complex and multifaceted construct that has been the subject of extensive research across various educational contexts. The literature identifies numerous factors that can influence a learner's satisfaction with their educational experience. We broadly categorize the features related to learning satisfaction into individual learner characteristics, instructional design elements, and social and environmental factors. The major learning satisfaction features of our work can be found in Table 1.

*2.2.1 Individual Learner Characteristics.*

Several studies have highlighted the importance of individual learner characteristics in determining their satisfaction. For instance, Artino and Stephens [5] suggest that learners' self-efficacy, or belief in their ability to succeed in specific tasks, can significantly impact their satisfaction with learning experiences. Similarly, using multiple regression analysis, Shen et al. [6] demonstrated that an individual's perception or belief in their ability to successfully complete an online course is a major determinant of satisfaction in online learning. Additionally, demographic variables such as age, gender, and prior educational background have been examined, with mixed findings on their influence on learning satisfaction ([7], [8]).

*2.2.2 Instructional Design Elements*

The quality and design of instructional materials are critical factors affecting learning satisfaction**.** Clear objectives, engaging content, and appropriate assessment methods contribute to higher satisfaction levels [10]. When it comes to online learning satisfaction, Costley and Lange [11] adopted a quasi-experimental design to investigate the effects of instructor-control on learners' online learning satisfaction and found that instructor control of learning environments though instructional design could positively affect learners' perceived learning satisfaction. This is strengthened by Yu's [8] finding that instructors' online teaching ability is the primary factor that affects learners' online learning satisfaction.

*2.2.3 Social and Environmental Factors*

The social context of learning, including interactions with peers and instructors, significantly influences learning satisfaction. According to Moore [12], the interaction between instructors and learners, including timely feedback and support, plays a vital role in fostering a satisfying learning environment. Rovai [13] believes that interaction among the learners is the key element of classroom community since learning occurs through the learners' active participation in class activities. With regards to online learning satisfaction, it has been found that learner-learner interaction ([14], [15]), learner-instructor interaction ([16], [17]), and the learner's interaction with content ([18], [19]) affect online learning satisfaction. The physical or virtual learning environment, including classroom design and online platform usability, also affects learners' satisfaction[20]. With the rise of e-learning, technology has become an increasingly important factor in learning satisfaction, especially after the COVID-19 pandemic. Technical issues and the ease of use of learning management systems can either facilitate or hinder satisfaction **Error! Reference source not found.**.

# 3 SURVEY DESIGN

Our challenge was to measure learning satisfaction to capture its multifaceted nature and to reflect the unique circumstances presented by the COVID-19 pandemic. Based on the literature review above, we define learning satisfaction as *the level of fulfillment an individual experiences with their educational experience*. We also categorize the factors of learning satisfaction into individual learner characteristics, instructional design elements, social and environmental factors as stated in the previous section.

An online survey containing questions on students' learning experiences, emotional state, stress management, and online learning environment during 2020 and 2021 was conducted. Out of the 302 participants, 88% were full-time students in South Korea and 12% were exchange students, with a balanced mix of genders and majors. Responses ranged from 5-point Likert-type values for opinion-based questions ("Strongly Agree" to "Strongly Disagree") to binary ("Yes"/"No") to one-of-*N* answers for factual information. Pre-processing included grouping students' majors into STEM, HASS, or hybrid categories, and manually checking text-supplied majors against the university's list. Course codes were similarly categorized. Questions were encoded into simpler column names for easier processing, and Likert responses were given numerical values for conducting Wilcoxon Rank Sum test and EBM model analysis. Survey distribution methods included course announcements, advertisements, and direct contact. Selected survey questions are provided in Table 1.

| Category | Survey Question | Encoding |
|---|---|---|
| Individual Learner's Characteristics | I feel comfortable with the way this course is conducted. | m_comfort |
| | I find this course boring. | m_boring |
| | I believe the things we studied in this course could be of some value to me. | m_valuable |
| | I think I am doing pretty good in this course. | m_selfEval |
| | I am satisfied with my performance at the tasks given in the lessons. | m_taskSatisfaction |
| | I think that I could be helpful to my classmates. | m_helpful |
| | I am able to have discussions with other classmates easily. | m_discuss |
| | I think that participating in class activities could help me to develop myself. | m_activity |
| | I feel I have been missing out on proper learning. | emo_miss |
| | Learner's First Major. | fstMajor |
| Instructional Design Elements | The learning method is suitable for this course. | m_suitable |
| | I get support from the teaching staff when I have trouble with the course. | m_ta |
| | I am sometimes frustrated because I cannot get instant feedback. | m_feedback |

| Category | Survey Question | Encoding |
|---|---|---|
| Social and Environmental Factors | I have lost friendships/relationships during this period. | emo_relationship |
| | I do creative things like art, writing, composing music, etc. to relieve stress. | cop_creative |
| | I do rigorous activities like high energy sports to relieve stress. | cop_sports |
| | I normally study with one or two friends during the pandemic. | env_group |
| | I have sufficient technological tools, to help with online learning. | env_tool |

Table 1: The main features and corresponding survey questions related to learning satisfaction, and their encodings.[1]

## 4 SIGNIFICANT COMPARISONS OF LEARNING ATTRIBUTES BETWEEN GROUPS

We found significant differences between the medians of two distinct paired groups of students in some of the major attributes related to learning satisfaction. Firstly, we found differences between groups that took online as opposed to offline courses, and secondly, between groups that majored in STEM and HASS. Each comparison was done using a two-sided Wilcoxon Sum Rank Test with 95% confidence interval. Table 2 shows that four attributes and seven attributes are significantly different between the online-offline and STEM-HASS groups respectively. The next two sections will explain these differences in more detail.

| Encoding | Online/Offline | STEM/HASS |
|---|---|---|
| m_comfort | Not significant | 0.0026090 |
| m_suitable | 0.0004742 | 2.957e-05 |
| m_valuable | Not significant | 0.0008396 |
| m_taskSatisfaction | Not significant | 0.0034280 |
| m_helpful | Not significant | 0.0205000 |
| m_concentrate | Not significant | 0.0059710 |
| m_feedback | 0.0005300 | Not significant |
| emo_miss | 2.612e-08 | Not significant |
| m_boring | 0.0016910 | 0.0011720 |

Table 2: Summary of Wilcoxon Sum Rank Test ($p$-values) for online vs. offline students and STEM vs. HASS students across nine learning attributes.

### 4.1 Comparison Between Online and Offline (In-class) Courses

Several studies have explored learners' satisfaction with online learning during the pandemic. For instance, Alqurshi [21] and Xhelili et al. [22] found that despite appreciating online learning's flexibility, students often grapple with technological issues, lack of motivation, and lack of face-to-face interaction, which significantly affects their overall satisfaction. In a similar vein, we discovered that online learning satisfaction was significantly lower in terms of student interaction and motivation, as indicated by the metrics *m_feedback* and *m_boring*.

Students who took offline courses were generally more satisfied, as can be seen by the longer blue bar for the positive question and shorter gray bars for the negative questions in Fig. 1. The longer red bars for online methods shows that students who took online courses were generally more dissatisfied and less motivated than students who took offline courses. On average, they felt they were missing out on proper learning 1.7 times more than students who took offline courses. They were also around 3.6 times more frustrated than their offline counterparts that they could not get instant feedback. Finally, students who took online courses also found their courses more boring than those who took offline courses by a factor of 1.8. These results clearly show that offline learning is superior to online learning when it comes to learning satisfaction and motivation. Nonetheless, we do have to acknowledge that students who took offline courses also felt that they have been slightly missing out on proper learning, did not get instant feedback and found their courses boring during the pandemic.

---

[1] The entire survey is available at *https://forms.gle/3bEMbszuUDW3MnQx5*

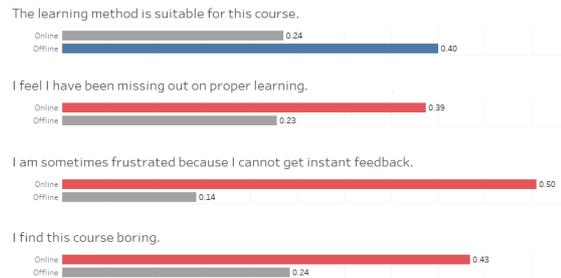

Fig. 1: Students' opinions on a course's learning method based on the mode of learning. The scores represent the median level of agreement for each question.

### 4.2 Comparison Between STEM and HASS students

Researchers are divided on the extent to which demographic factors such as gender and age impact satisfaction with online learning [8]. Shen et al. [6] surveyed 406 online students enrolled in an online course and found that female students experienced higher levels of online learning satisfaction than male ones. However, through quantitative analysis of data collected from 392 students enrolled in 28 online courses, Ke and Kwak [9] found that online learners' age didn't influence learners' satisfaction. Hettiarachchi et al. [23] also found no significant influence of gender on students' online learning satisfaction.

Similar to the latter findings above, our study also found that demographic factors such as age do not pose any significant impact on online learning satisfaction. However, there are significant differences between the learning preferences of different study majors when it comes to learning satisfaction. The bars in Fig. 2 highlight these differences. STEM students, on average, had higher satisfaction levels than HASS students in these seven attributes. In the top five bars, the median of STEM students' satisfaction levels are 1.3 to 1.7 times higher than HASS students.

The differences in the degree of learning satisfaction between the two groups could be attributed to several reasons. We keep our discussion centred on the assumption that most of the learning was done online during this period. The first one is the difference in familiarity with technology. A student's adaptation depends on the level of awareness, the level of familiarity with information technology, and the willingness to get involved and adapt to the e-learning approach [24]. STEM majors typically require more use of technology in their coursework, and this intuitively implies that they are more comfortable with online learning methods. In contrast, HASS majors may be less familiar with technology or may not have had as much exposure to online learning methods, which could make the transition more challenging. A recent study [25] revealed that STEM students have better digital profiles before the pandemic than non-STEM students, enabling them to cope with online learning during the pandemic more quickly than non-STEM students. Another reason could be the difference in learning styles of the courses in these majors. The nature of the course content could also play a role in students' satisfaction with online learning. STEM courses may be more focused on facts, theories, and problem-solving, which may be easier to teach and learn online. Chirikov et al. [26] found that STEM students learn just as much in online classrooms as they do in offline ones. HASS courses, on the other hand, may be more focused on critical thinking, interpretation, and discussion, which may be harder to replicate in an online environment.

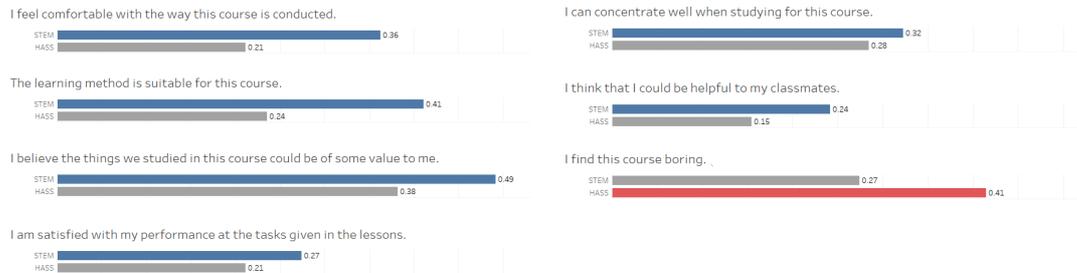

Fig. 2: Students' opinions on a course's learning method based on their first major. The scores represent the median level of agreement for each question.

The key takeaway is that students who took offline courses during the pandemic had higher learning satisfaction, among other attributes, than those who took online courses. The same could be said about STEM students as compared to their HASS counterparts. Seeing that learning satisfaction is the most prominent attribute differentiating these two groups of students, we probe into the factors that influence students' learning satisfaction and unpack the interpretations for them in the following section.

## 5 FACTORS AFFECTING STUDENTS' LEARNING SATISFACTION

In this section, we employ an interpretable machine learning model to determine the significant factors that affect students' learning satisfaction. This complements and extends the findings of the previous section. We provide an overview of the task and model, followed by the experimental setup and interpretation of the results yielded by the model.

### 5.1 Task Design

Our study involved a binary classification task to investigate factors impacting student satisfaction during the pandemic. The task used multiple input features to predict the target outcome, students' learning satisfaction. This target was represented using an OR combination of two features: the perceived value of the studied material ($m\_valuable$) and satisfaction with task performance ($m\_taskSatisfaction$). These were chosen as they best reflect student's learning satisfaction, in line with our definition in Section 3. Responses, ranging from 'Strongly Agree' to 'Strongly Disagree', were coded as either positive (satisfied) or negative (not satisfied).

### 5.2 Model Overview

The Explainable Boosting Machine (EBM) [27] is a machine learning model that balances performance and interpretability. It's an enhanced Generalized Additive Model ($GAM$) that allows users to understand how individual input features contribute to the model. $GAM$ fits arbitrary functions into a generalized linear model to find nonlinear relationships between targets and inputs. The target, $y$, is expressed as a combination of arbitrary functions of its inputs, $x_i$, as shown in Equation 1:

$$g(E[y]) = \Sigma f_i(x_i) \quad (1)$$

Here, $g$ is the link function that modifies the GAM for regression or classification, and $f_i$ is a shape function for each feature. For binary classification, $g$ is the logistic function $\frac{1}{1+e^{-E[y]}}$. EBM is further evolved into a Generalized Additive Model plus Interaction ($GA^2M$), as shown in Equation 2 below:

$$g(E[y]) = \Sigma f_i(x_i) + \Sigma f_{ij}(x_i, x_j) \quad (2)$$

This model considers both univariate inputs and pairwise interactions of two inputs with the target. This allows it to interpret how individual features and interactions between two inputs affect the target variable. Feature selection is discussed in the following section.

### 5.3 Data Distribution and Feature Selection

Survey data was split into training and test sets in an 80/20 ratio with 241 and 61 samples respectively. The data was imbalanced with the positive class being dominant. To create more balance in the dataset, we used Synthetic Minority Over-sampling Technique (SMOTE) [28]

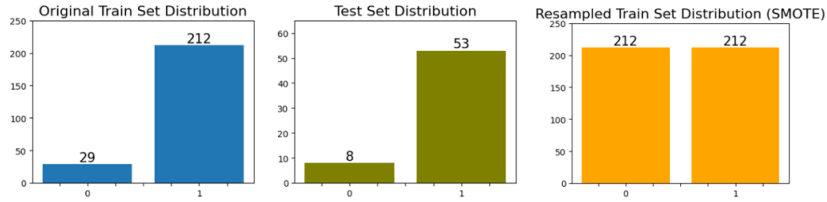

Fig. 3: Data distribution according to positive and negative samples for train/test sets, and train set after SMOTE resampling.

### 5.4 Results

The EBM model, fitted to the dataset for learning satisfaction, yielded an accuracy of 95.08% and an F1 score of 94.33, suggesting reliable features impacting student learning satisfaction. The F1 score, combining precision and recall, offers a comprehensive evaluation. These features, detailed in Table 3, are ranked, and scored for their contribution to learning satisfaction in the following subsection.

| Accuracy | F1 Score |
|---|---|
| 95.08 | 94.33 |

Table 3: EBM model performance for learning satisfaction (*m_valuable* OR *m_taskSatisfaction*).

*5.4.1 Feature Ranking*

Fig. 4 displays the feature importance ranking in our EBM model, determined by the mean absolute score. This score, referred to as feature importance score (FIS), is the weighted average absolute value calculated for each feature across the training dataset. Essentially, it measures the average impact of each feature on the model's decision-making.

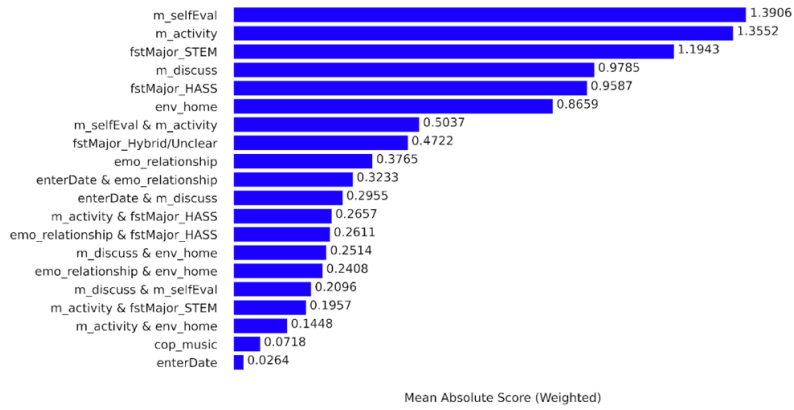

Fig. 4: Ranking of the top 20 features in order of their contributions to the model's decision making.

The primary determinants of students' learning satisfaction are their self-perceived performance (*m_selfEval, FIS=1.3906*), engagement in class activities (*m_activity, FIS=1.3552*), and having a STEM major (*fstMajor_STEM, FIS=1.1943*), as per the EBM model. The model considers both individual and paired features, with the latter revealing interactions between features. The impact of these features on learning satisfaction varies with their value changes.

*5.4.2 Single Feature Importance*

Fig. 5's plots summarize the relationship between individual inputs and learning satisfaction. Features on the *x*-axis range from 0 to 1 for binomial features, and -1 to 1 for opinion-based features. The *y*-axis represents predictive scores, with higher scores indicating a higher likelihood of positive class occurrence. We begin with weakly related features, followed by positively and negatively related ones.

Fig. 5(a) suggests similar learning satisfaction for students who entered university before and during the pandemic, with a minor score difference of 0.55. The low FIS of 0.03 means that *enter_date* has less impact on learning satisfaction than other features like *m_activity* and *m_selfEval*. Fig. 5(b) shows that students with positive self-evaluation are more satisfied with their learning (score=2.091), whereas indifferent or self-perceived poor performers are less satisfied. Similarly, Fig 5(c) demonstrates higher learning satisfaction among students who can easily engage in class discussions. Thus, perceived performance and class engagement levels significantly influence learning satisfaction.

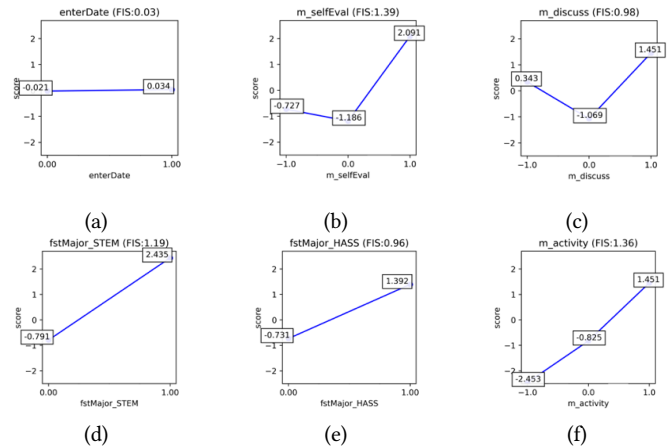

Fig. 5: Visualizations of the scores for selected single input features used by the EBM model

When considering students' first major, both STEM and HASS students were generally satisfied, with STEM students demonstrating higher satisfaction (score=2.435) than HASS students (score=1.392). The first major being STEM was the third most significant factor in the model's decision-making process (FIS=1.19). We also evaluated students' perceptions of class participation. Students who saw participation as beneficial were satisfied (score=1.451), while those who did not were generally dissatisfied (score=-2.453). This feature had a significant impact on learning dissatisfaction and was the second most influential factor in determining learning satisfaction.

*5.4.3 Pairwise Feature Importance*
We examined the impact of feature interactions on learning satisfaction by creating heatmaps of paired features, revealing their combined effect. Positive and negative pairwise scores indicate positive and negative impacts, respectively. The feature values and scale remain consistent with prior sections, allowing comparison between single feature plots and these pairwise ones.

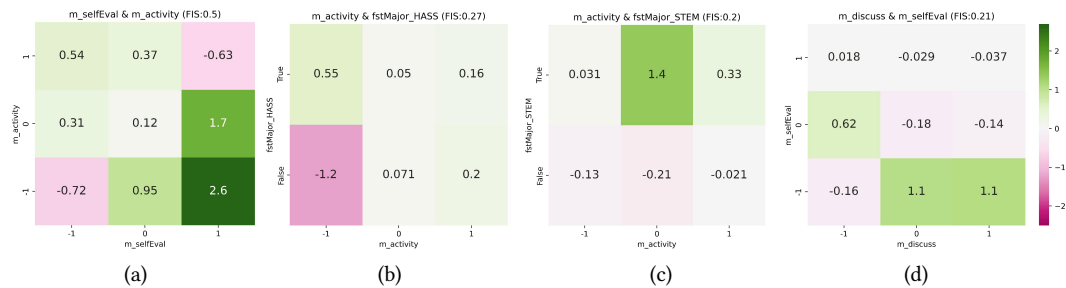

Fig. 6: Heatmaps for selected top contributing pairwise feature interactions.

In the previous section, positive perceived performance contributed to positive learning satisfaction while negative student engagement led to negative learning satisfaction. We will now see how the combination of both these features influenced learning satisfaction, termed *m_selfEval & m_activity*. Fig 6(a) shows that students with high self-perceived performance but who found class activities unhelpful *(m_selfEval=1 & m_activity= -1)* still had high learning satisfaction (pairwise score=2.6), the highest of all paired features and the most influential on model's decision making (FIS=0.5, ranked 7th).

Students who perceived themselves as underperforming and found class participation unhelpful were generally dissatisfied with their learning (*m_selfEval= -1 & m_activity= -1,* pairwise score=-0.72). This likely reflects low performers who found class discussions insufficient, especially during the pandemic. Interestingly, students with high perceived performance were generally satisfied except when they found class activities beneficial (pairwise score=-0.63). This counterintuitive finding may pertain to students who perceived themselves as performing well but had actual low performance. Even if they found class discussions useful, they remained dissatisfied with their overall learning.

The interaction between student engagement and study majors reveals that students who found class activities unhelpful and were not HASS majors were the least satisfied group, with a negative pairwise score of -1.2 (Fig. 6(b)). This dissatisfaction is greater than those who were not satisfied with both their performance and class activities. HASS students were slightly satisfied even if they found class activities unhelpful (pairwise score=0.55). STEM students, however, were generally satisfied regardless of their views on class activities, especially those indifferent to participation (Fig. 6(c), pairwise score=1.4).

Fig 6(d) indicates that students, not restricted from peer discussions and with low self-perceived performance *(m_discuss > -1 & m_selfEval = -1)*, were generally content with their learning (pairwise score=1.1). Comparatively, some students who struggled with study space and self-perception managed to offset these by discussing with peers. Those who had these discussions comfortably were content. Even those who couldn't easily discuss but didn't perceive low performance *(m_discuss= -1 & m_selfEval > -1)* were satisfied. This suggests that students with medium or high self-perception found alternative ways to compensate for classroom communication deficits, enhancing their learning satisfaction.

## 6 DISCUSSION AND FUTURE OUTLOOK

Our research analysed factors affecting student learning satisfaction during the pandemic in 302 students in South Korea. Statistical tests have revealed that students' learning satisfaction is higher for those who take offline classes as compared to online classes and higher for STEM students than their HASS counterparts. Explainable AI has shown that students' perceived performance, their first major being STEM, their perception on participating in class activities and their ability to discuss with their peers are the main factors that influence their learning satisfaction. Delving deeper, we saw that positive perceived performance and the ability to discuss with classmates had positive impact on learning satisfaction, as well as their first major being STEM. Negative perception on class activities participation had negative impact on learning satisfaction. However, when assessed in combination, students with high perceived performance who did not find class activities helpful were still satisfied with their learning. Another discovery is that STEM students were generally satisfied with their learning regardless of whether they found participating in class activities helpful or not, while HASS students were satisfied with their learning if they found participating in class activities were helpful. Our findings could aid institutions and educators undertake improved educational strategies from understanding the profiles of certain student groups. We would like to extend our work by examining other crucial aspects such as students' perceived performance and motivation. Additionally, we aim to use an explainable deep learning model for further analysis and comparison of results.